\documentclass[prb,preprintnumbers,amsmath,amssymb,floatfix]{revtex4}

\usepackage{graphicx}
\usepackage{subfigure}
\usepackage{dcolumn}

\large
\begin{document}

\title{Spinor wave equation of photon}
\author{Xiang-Yao Wu$^{a}$ \footnote{E-mail: wuxy2066@163.com},
Bo-Jun Zhang$^{a}$, Xiao-Jing Liu$^{a}$, Si-Qi Zhang$^{a}$, Jing
Wang$^{a}$, Hong Li$^{a}$, Xi-Hui Fan$^{b}$ and  Jing-Wu Li$^{c}$}
\affiliation{a. Institute of Physics, Jilin Normal University,
Siping 136000 China\\
b. Institute of Physics, Qufu Normal University, Xuzhou 273165
China\\
c. Institute of Physics, Xuzhou Normal University, Xuzhou 221000,
China}

\begin{abstract}
In this paper, we give the spinor wave equations of free and
unfree photon, which are the differential equation of space-time
one order. For the free photon, the spinor wave equations are
covariant, and the spinors $\psi$ are corresponding to the the
reducibility representations $D^{10}+D^{01}$ and
$D^{10}+D^{01}+D^{\frac{1}{2}\frac{1}{2}}$ of the proper Lorentz
group.
\\
\vskip 5pt

PACS: 03.65.-w, 03.50.De, 42.50.Xa \\
Keywords: Quantum theory; Spinor equation; Photon;

\end{abstract}
\maketitle

{\bf 1. Introduction} \vskip 8pt

Photons are quantum particles that their behavior is governed by
the laws of quantum mechanics. This means, their state are
described by wave functions. According to modern quantum field
theory, photons, together with all other particles, are the
quantum excitations of a field. In the case of photons, these are
the excitations of the electromagnetic field. The lowest field
excitation of a given type corresponds to one photon and higher
field excitations involve more than one photon. This concept of a
photon enables one to use the photon wave function not only to
describe quantum states of an excitation of the free field but
also of the electromagnetic field interacting with a medium.
Maxwell equations in the matrix Dirac-like form considered during
long time by many authors, the interest to the
Majorana-Oppenheimer formulation of electrodynamics has grown in
recent years [1-14].

After discovering the relativistic equation for a particle with
spin 1/2 [15], In Refs. [1-3], the author have proposed to
consider the Maxwell theory of electromagnetism as the wave
mechanics of the photon, then it must be possible to write Maxwell
equations as a Dirac-like equation for a probability quantum wave
$\vec{\psi}$, this wave function being expressable by means of the
physical $E$, $B$ fields, and the complex 3-vector wave function
satisfying the massless Dirac-like equations. Afterwards, much
work was done to study spinor and vectors within the Lorentz group
theory: Moglich [16], Ivanenko-Landau [17], Neumann [18], van der
Waerden [19]. As was shown any quantity which transforms linearly
under Lorentz transformations is a spinor. For that reason spinor
quantities are considered as fundamental in quantum field theory
and basic equations for such quantities should be written in a
spinor form. A spinor formulation of Maxwell equations was studied
by Laporte and Uhlenbeck [20-22]. In this paper, we give the
spinor wave equations of free and unfree photon, which are the
differential equation of space-time one order. For the free
photon, the spinor wave equations are covariant, and the spinors
$\psi$ are corresponding to the the reducibility representations
of the proper Lorentz group.

\vskip 8pt

{\bf 2. Spinor wave equation of free photon}

\vskip 8pt

The differential equation of space-time two order for free
electromagnetism wave is
\begin{equation}
\Box A_{\mu}=0,
\end{equation}
and the Lorentz condition is
\begin{equation}
\partial_{\mu}A_{\mu}=0,
\end{equation}
where
\begin{equation}
\partial_{\mu}=(\nabla,\frac{\partial}{\partial(ict)}),\hspace{0.5in}
\Box=\partial_{\mu}\partial_{\mu}=\nabla^{2}-\frac{1}{{c^{2}}}
\frac{1}{\partial t^{2}}.
\end{equation}
The equation (1) has four components $A_{\mu}  (\mu=1, 2, 3, 4)$.
making Eq. (1) into one order differential equation of space-time,
the number of field functions should be added.\\
Defining the function
\begin{equation}
F_{\nu\mu}=\partial_{\nu}A_{\mu},
\end{equation}
and Eq. (1) becomes
\begin{eqnarray}
\partial_{\nu}F_{\nu\mu}=\partial_{\nu}\partial_{\nu}A_{\mu}=\Box
A_{\mu}=0,
\end{eqnarray}
i.e., the differential equation of space-time one order for free
electromagnetism wave is
\begin{eqnarray}
\partial_{\nu}F_{\nu\mu}=0.
\end{eqnarray}
By Eqs. (2) and (4), we have
\begin{equation}
\partial_{\mu}F_{\nu\mu}=\partial_{\nu}\partial_{\mu}A_{\mu}=0,
\end{equation}
with Eqs. (6) and (7), there is
\begin{eqnarray}
\partial_{\nu}F_{\nu\mu}=\partial_{\mu}F_{\nu\mu}=\partial_{\nu}F_{\mu\nu}=0.
\end{eqnarray}
From Eq. (8), we have
\begin{eqnarray}
F_{\mu\nu}=F_{\nu\mu},
\end{eqnarray}
or
\begin{eqnarray}
F_{\mu\nu}=-F_{\nu\mu}.
\end{eqnarray}
In Eq. (9), $F_{\mu\nu}$ is symmetry tensor, it has ten
independent components, and its matrix is
\begin{eqnarray}
F_{\mu\nu}= \left ( \begin{array}{llll}
   F_{11}  \hspace{0.3in} F_{12}  \hspace{0.3in} F_{13}  \hspace{0.3in} F_{14}\\
   F_{12}  \hspace{0.3in} F_{22}  \hspace{0.3in} F_{23}  \hspace{0.3in} F_{24}\\
   F_{13}  \hspace{0.3in} F_{23}  \hspace{0.3in} F_{33}  \hspace{0.3in} F_{34}\\
   F_{14}  \hspace{0.3in} F_{24}  \hspace{0.3in} F_{34}  \hspace{0.3in} F_{44}\\
   \end{array}
   \right ),
\end{eqnarray}
In Eq. (10), $F_{\mu\nu}$ is antisymmetry tensor, it has six
independent components, and its matrix is
\begin{eqnarray}
F_{\mu\nu}= \left ( \begin{array}{llll}
\hspace{0.15in}0   \hspace{0.3in} F_{12}  \hspace{0.25in} F_{13}  \hspace{0.25in} F_{14}\\
   -F_{12}   \hspace{0.3in} 0  \hspace{0.3in} F_{23}  \hspace{0.25in} F_{24}\\
   -F_{13}  \hspace{0.1in} -F_{23}  \hspace{0.3in} 0  \hspace{0.25in} F_{34}\\
   -F_{14}  \hspace{0.1in} -F_{24}  \hspace{0.1in} -F_{34}  \hspace{0.25in} 0\\
   \end{array}
   \right ).
\end{eqnarray}
For the proper Lorentz group $L_{p}$, the irreducibility
representations of spin $1$ particle or field are $D^{10}$,
$D^{01}$ and $D^{\frac{1}{2}\frac{1}{2}}$, respectively, and the
dimension of irreducibility representations are corresponding to
three, three and four. Therefore, the reducibility representations
of particle or field with spin $1$ are
\begin{equation}
D=D^{10}+ D^{01},
\end{equation}
\begin{equation}
D=D^{10}+ D^{01}+ D^{\frac{1}{2}\frac{1}{2}},
\end{equation}
\begin{eqnarray}
\cdot\cdot\cdot\nonumber
\end{eqnarray}
Eqs. (13) and (14) are corresponding to six and ten dimensions
irreducibility representations, which are the two lowest
dimensional irreducibility representations

When $F_{\mu\nu}$ take the antisymmetry tensor of Eq. (12), which
is the representation vector of reducibility representations
$D=D^{10}+ D^{01}$. Eq. (6) can be written as
\begin{eqnarray}
&& \mu=1:
\hspace{0.1in}\partial_{1}F_{11}+\partial_{2}F_{21}+\partial_{3}F_{31}+\partial_{4}F_{41}=0,\\&&
\mu=2:
\hspace{0.1in}\partial_{1}F_{12}+\partial_{2}F_{22}+\partial_{3}F_{32}+\partial_{4}F_{42}=0,
\\&& \mu=3:
\hspace{0.1in}\partial_{1}F_{13}+\partial_{2}F_{23}+\partial_{3}F_{33}+\partial_{4}F_{43}=0,
\\&& \mu=4:
\hspace{0.1in}\partial_{1}F_{14}+\partial_{2}F_{24}+\partial_{3}F_{34}+\partial_{4}F_{44}=0,
\end{eqnarray}
substituting Eq. (12) into (15)-(18), there is
\begin{eqnarray}
\left \{ \begin{array}{lll} &&
\partial_{2}F_{12}+\partial_{3}F_{13}+\partial_{4}F_{14}=0\\&&
\partial_{1}F_{12}-\partial_{3}F_{23}-\partial_{4}F_{24}=0
\\&& \partial_{1}F_{13}+\partial_{2}F_{23}-\partial_{4}F_{34}=0
\\&& \partial_{1}F_{14}+\partial_{2}F_{24}+\partial_{3}F_{34}=0
    \end{array}
   \right..
\end{eqnarray}
Eq. (19) can be written as the differential form of space-time one
order
\begin{eqnarray}
\beta_{\mu}\partial_{\mu}\psi=0 ,\hspace{0.1in} (\mu=1, 2, 3, 4),
\end{eqnarray}
where the spinor wave function $\psi$ is
\begin{eqnarray}
\psi= \left ( \begin{array}{llllll}
 F_{12}\\
 F_{13}\\
 F_{14}\\
 F_{23}\\
 F_{24}\\
 F_{34}\\
   \end{array}
   \right )=\left ( \begin{array}{llllll}
 \psi_{1}\\
 \psi_{2}\\
 \psi_{3}\\
 \psi_{4}\\
 \psi_{5}\\
 \psi_{6}\\
   \end{array}
   \right ),
\end{eqnarray}
and the $\beta$ matrixes are
\begin{eqnarray}
&&\beta_{1}= \left ( \begin{array}{llllll}
 \hspace{0.2in}0   \hspace{0.2in} 0  \hspace{0.2in} 0  \hspace{0.2in} 0 \hspace{0.2in} 0  \hspace{0.2in} 0\\
 \hspace{0.2in}1   \hspace{0.2in} 0  \hspace{0.2in} 0  \hspace{0.2in} 0 \hspace{0.2in} 0  \hspace{0.2in} 0\\
 \hspace{0.2in}0   \hspace{0.2in} 1  \hspace{0.2in} 0  \hspace{0.2in} 0 \hspace{0.2in} 0  \hspace{0.2in} 0\\
 \hspace{0.2in}0   \hspace{0.2in} 0  \hspace{0.2in} 1  \hspace{0.2in} 0 \hspace{0.2in} 0  \hspace{0.2in} 0\\
 \hspace{0.2in}0   \hspace{0.2in} 0  \hspace{0.2in} 0  \hspace{0.2in} 0 \hspace{0.2in} 0  \hspace{0.2in} 0\\
 \hspace{0.2in}0   \hspace{0.2in} 0  \hspace{0.2in} 0  \hspace{0.2in} 0 \hspace{0.2in} 0  \hspace{0.2in} 0\\
   \end{array}
   \right ),\beta_{2}= \left ( \begin{array}{llllll}
 \hspace{0.2in}1   \hspace{0.2in} 0  \hspace{0.2in} 0  \hspace{0.2in} 0 \hspace{0.2in} 0  \hspace{0.2in} 0\\
 \hspace{0.2in}0   \hspace{0.2in} 0  \hspace{0.2in} 0  \hspace{0.2in} 0 \hspace{0.2in} 0  \hspace{0.2in} 0\\
 \hspace{0.2in}0   \hspace{0.2in} 0  \hspace{0.2in} 0  \hspace{0.2in} 1 \hspace{0.2in} 0  \hspace{0.2in} 0\\
 \hspace{0.2in}0   \hspace{0.2in} 0  \hspace{0.2in} 0  \hspace{0.2in} 0 \hspace{0.2in} 1  \hspace{0.2in} 0\\
 \hspace{0.2in}0   \hspace{0.2in} 0  \hspace{0.2in} 0  \hspace{0.2in} 0 \hspace{0.2in} 0  \hspace{0.2in} 0\\
 \hspace{0.2in}0   \hspace{0.2in} 0  \hspace{0.2in} 0  \hspace{0.2in} 0 \hspace{0.2in} 0  \hspace{0.2in} 0\\
   \end{array}
   \right )
\nonumber\\&&\beta_{3}= \left ( \begin{array}{llllll}
 \hspace{0.2in}0   \hspace{0.2in} 1  \hspace{0.2in} 0  \hspace{0.2in} 0 \hspace{0.2in} 0  \hspace{0.2in} 0\\
 \hspace{0.2in}0   \hspace{0.2in} 0  \hspace{0.2in} 0  \hspace{0.05in} -1 \hspace{0.18in} 0  \hspace{0.2in} 0\\
 \hspace{0.2in}0   \hspace{0.2in} 0  \hspace{0.2in} 0  \hspace{0.2in} 0 \hspace{0.2in} 0  \hspace{0.2in} 0\\
 \hspace{0.2in}0   \hspace{0.2in} 0  \hspace{0.2in} 0  \hspace{0.2in} 0 \hspace{0.2in} 0  \hspace{0.2in} 1\\
 \hspace{0.2in}0   \hspace{0.2in} 0  \hspace{0.2in} 0  \hspace{0.2in} 0 \hspace{0.2in} 0  \hspace{0.2in} 0\\
 \hspace{0.2in}0   \hspace{0.2in} 0  \hspace{0.2in} 0  \hspace{0.2in} 0 \hspace{0.2in} 0  \hspace{0.2in} 0\\
   \end{array}
   \right ),\beta_{4}= \left ( \begin{array}{llllll}
 \hspace{0.2in}0   \hspace{0.2in} 0  \hspace{0.2in} 1  \hspace{0.2in} 0 \hspace{0.2in} 0  \hspace{0.2in} 0\\
 \hspace{0.2in}0   \hspace{0.2in} 0  \hspace{0.2in} 0  \hspace{0.2in} 0 \hspace{0.1in} -1  \hspace{0.13in} 0\\
 \hspace{0.2in}0   \hspace{0.2in} 0  \hspace{0.2in} 0  \hspace{0.2in} 1 \hspace{0.2in} 0  \hspace{0.1in} -1\\
 \hspace{0.2in}0   \hspace{0.2in} 0  \hspace{0.2in} 0  \hspace{0.2in} 0 \hspace{0.2in} 0  \hspace{0.2in} 0\\
 \hspace{0.2in}0   \hspace{0.2in} 0  \hspace{0.2in} 0  \hspace{0.2in} 0 \hspace{0.2in} 0  \hspace{0.2in} 0\\
 \hspace{0.2in}0   \hspace{0.2in} 0  \hspace{0.2in} 0  \hspace{0.2in} 0 \hspace{0.2in} 0  \hspace{0.2in} 0\\
   \end{array}
   \right ).
\end{eqnarray}
When $F_{\mu\nu}$ take the symmetry tensor of Eq. (11), which is
the representation vector of reducibility representations
$D=D^{10}+D^{01}+D^{\frac{1}{2}\frac{1}{2}}$. Eq. (6) can be
written as
 \begin{eqnarray}
\left \{ \begin{array}{llll} &&
\partial_{1}F_{11}+\partial_{2}F_{12}+\partial_{3}F_{13}+\partial_{4}F_{14}=0\\&&
\partial_{1}F_{12}+\partial_{2}F_{22}+\partial_{3}F_{23}+\partial_{4}F_{24}=0\\
&&\partial_{1}F_{13}+\partial_{2}F_{23}+\partial_{3}F_{33}+\partial_{4}F_{34}=0\\&&
\partial_{1}F_{14}+\partial_{2}F_{24}+\partial_{3}F_{34}+\partial_{4}F_{44}=0
    \end{array}
   \right..
\end{eqnarray}
Eq. (23) can be written as the differential form of space-time one
order
\begin{equation}
\beta_{\mu}\partial_{\mu}\psi=0,
\end{equation}
where the spinor wave function $\psi$ is
\begin{eqnarray}
\psi=\left (
\begin{array}{llllllllll}
  F_{11}\\
  F_{12}\\
  F_{13}\\
  F_{14}\\
  F_{22}\\
  F_{23}\\
  F_{24}\\
  F_{33}\\
  F_{34}\\
  F_{44}\\
  \end{array}
  \right )=\left (
\begin{array}{llllllllll}
  \psi_{1}\\
  \psi_{2}\\
  \psi_{3}\\
  \psi_{4}\\
  \psi_{5}\\
  \psi_{6}\\
  \psi_{7}\\
  \psi_{8}\\
  \psi_{9}\\
  \psi_{10}\\
  \end{array}
  \right ),
\end{eqnarray}
and the $\beta$ matrixes are
\begin{eqnarray}
\beta_{1}= \left ( \begin{array}{llllllllll}
   1  \hspace{0.3in} 0 \hspace{0.3in} 0 \hspace{0.3in} 0 \hspace{0.3in} 0 \hspace{0.3in} 0 \hspace{0.3in} 0 \hspace{0.3in} 0 \hspace{0.3in} 0 \hspace{0.3in} 0\\
   0  \hspace{0.3in} 1 \hspace{0.3in} 0 \hspace{0.3in} 0 \hspace{0.3in} 0 \hspace{0.3in} 0 \hspace{0.3in} 0 \hspace{0.3in} 0 \hspace{0.3in} 0 \hspace{0.3in} 0\\
   0  \hspace{0.3in} 0 \hspace{0.3in} 1 \hspace{0.3in} 0 \hspace{0.3in} 0 \hspace{0.3in} 0 \hspace{0.3in} 0 \hspace{0.3in} 0 \hspace{0.3in} 0 \hspace{0.3in} 0\\
   0  \hspace{0.3in} 0 \hspace{0.3in} 0 \hspace{0.3in} 1 \hspace{0.3in} 0 \hspace{0.3in} 0 \hspace{0.3in} 0 \hspace{0.3in} 0 \hspace{0.3in} 0 \hspace{0.3in} 0\\
   0  \hspace{0.3in} 0 \hspace{0.3in} 0 \hspace{0.3in} 0 \hspace{0.3in} 0 \hspace{0.3in} 0 \hspace{0.3in} 0 \hspace{0.3in} 0 \hspace{0.3in} 0 \hspace{0.3in} 0\\
   0  \hspace{0.3in} 0 \hspace{0.3in} 0 \hspace{0.3in} 0 \hspace{0.3in} 0 \hspace{0.3in} 0 \hspace{0.3in} 0 \hspace{0.3in} 0 \hspace{0.3in} 0 \hspace{0.3in} 0\\
   0  \hspace{0.3in} 0 \hspace{0.3in} 0 \hspace{0.3in} 0 \hspace{0.3in} 0 \hspace{0.3in} 0 \hspace{0.3in} 0 \hspace{0.3in} 0 \hspace{0.3in} 0 \hspace{0.3in} 0\\
   0  \hspace{0.3in} 0 \hspace{0.3in} 0 \hspace{0.3in} 0 \hspace{0.3in} 0 \hspace{0.3in} 0 \hspace{0.3in} 0 \hspace{0.3in} 0 \hspace{0.3in} 0 \hspace{0.3in} 0\\
   0  \hspace{0.3in} 0 \hspace{0.3in} 0 \hspace{0.3in} 0 \hspace{0.3in} 0 \hspace{0.3in} 0 \hspace{0.3in} 0 \hspace{0.3in} 0 \hspace{0.3in} 0 \hspace{0.3in} 0\\
   0  \hspace{0.3in} 0 \hspace{0.3in} 0 \hspace{0.3in} 0 \hspace{0.3in} 0 \hspace{0.3in} 0 \hspace{0.3in} 0 \hspace{0.3in} 0 \hspace{0.3in} 0 \hspace{0.3in} 0\nonumber\\
   \end{array}
   \right ),
\end{eqnarray}
\begin{eqnarray}
\beta_{2}= \left ( \begin{array}{llllllllll}
   0  \hspace{0.3in} 1 \hspace{0.3in} 0 \hspace{0.3in} 0 \hspace{0.3in} 0 \hspace{0.3in} 0 \hspace{0.3in} 0 \hspace{0.3in} 0 \hspace{0.3in} 0 \hspace{0.3in} 0\\
   0  \hspace{0.3in} 0 \hspace{0.3in} 0 \hspace{0.3in} 0 \hspace{0.3in} 1 \hspace{0.3in} 0 \hspace{0.3in} 0 \hspace{0.3in} 0 \hspace{0.3in} 0 \hspace{0.3in} 0\\
   0  \hspace{0.3in} 0 \hspace{0.3in} 0 \hspace{0.3in} 0 \hspace{0.3in} 0 \hspace{0.3in} 1 \hspace{0.3in} 0 \hspace{0.3in} 0 \hspace{0.3in} 0 \hspace{0.3in} 0\\
   0  \hspace{0.3in} 0 \hspace{0.3in} 0 \hspace{0.3in} 0 \hspace{0.3in} 0 \hspace{0.3in} 0 \hspace{0.3in} 1 \hspace{0.3in} 0 \hspace{0.3in} 0 \hspace{0.3in} 0\\
   0  \hspace{0.3in} 0 \hspace{0.3in} 0 \hspace{0.3in} 0 \hspace{0.3in} 0 \hspace{0.3in} 0 \hspace{0.3in} 0 \hspace{0.3in} 0 \hspace{0.3in} 0 \hspace{0.3in} 0\\
   0  \hspace{0.3in} 0 \hspace{0.3in} 0 \hspace{0.3in} 0 \hspace{0.3in} 0 \hspace{0.3in} 0 \hspace{0.3in} 0 \hspace{0.3in} 0 \hspace{0.3in} 0 \hspace{0.3in} 0\\
   0  \hspace{0.3in} 0 \hspace{0.3in} 0 \hspace{0.3in} 0 \hspace{0.3in} 0 \hspace{0.3in} 0 \hspace{0.3in} 0 \hspace{0.3in} 0 \hspace{0.3in} 0 \hspace{0.3in} 0\\
   0  \hspace{0.3in} 0 \hspace{0.3in} 0 \hspace{0.3in} 0 \hspace{0.3in} 0 \hspace{0.3in} 0 \hspace{0.3in} 0 \hspace{0.3in} 0 \hspace{0.3in} 0 \hspace{0.3in} 0\\
   0  \hspace{0.3in} 0 \hspace{0.3in} 0 \hspace{0.3in} 0 \hspace{0.3in} 0 \hspace{0.3in} 0 \hspace{0.3in} 0 \hspace{0.3in} 0 \hspace{0.3in} 0 \hspace{0.3in} 0\\
   0  \hspace{0.3in} 0 \hspace{0.3in} 0 \hspace{0.3in} 0 \hspace{0.3in} 0 \hspace{0.3in} 0 \hspace{0.3in} 0 \hspace{0.3in} 0 \hspace{0.3in} 0 \hspace{0.3in} 0\nonumber\\
   \end{array}
   \right ),
\end{eqnarray}
\begin{eqnarray}
\beta_{3}= \left ( \begin{array}{llllllllll}
   0  \hspace{0.3in} 0 \hspace{0.3in} 1 \hspace{0.3in} 0 \hspace{0.3in} 0 \hspace{0.3in} 0 \hspace{0.3in} 0 \hspace{0.3in} 0 \hspace{0.3in} 0 \hspace{0.3in} 0\\
   0  \hspace{0.3in} 0 \hspace{0.3in} 0 \hspace{0.3in} 0 \hspace{0.3in} 0 \hspace{0.3in} 1 \hspace{0.3in} 0 \hspace{0.3in} 0 \hspace{0.3in} 0 \hspace{0.3in} 0\\
   0  \hspace{0.3in} 0 \hspace{0.3in} 0 \hspace{0.3in} 0 \hspace{0.3in} 0 \hspace{0.3in} 0 \hspace{0.3in} 0 \hspace{0.3in} 1 \hspace{0.3in} 0 \hspace{0.3in} 0\\
   0  \hspace{0.3in} 0 \hspace{0.3in} 0 \hspace{0.3in} 0 \hspace{0.3in} 0 \hspace{0.3in} 0 \hspace{0.3in} 0 \hspace{0.3in} 0 \hspace{0.3in} 1 \hspace{0.3in} 0\\
   0  \hspace{0.3in} 0 \hspace{0.3in} 0 \hspace{0.3in} 0 \hspace{0.3in} 0 \hspace{0.3in} 0 \hspace{0.3in} 0 \hspace{0.3in} 0 \hspace{0.3in} 0 \hspace{0.3in} 0\\
   0  \hspace{0.3in} 0 \hspace{0.3in} 0 \hspace{0.3in} 0 \hspace{0.3in} 0 \hspace{0.3in} 0 \hspace{0.3in} 0 \hspace{0.3in} 0 \hspace{0.3in} 0 \hspace{0.3in} 0\\
   0  \hspace{0.3in} 0 \hspace{0.3in} 0 \hspace{0.3in} 0 \hspace{0.3in} 0 \hspace{0.3in} 0 \hspace{0.3in} 0 \hspace{0.3in} 0 \hspace{0.3in} 0 \hspace{0.3in} 0\\
   0  \hspace{0.3in} 0 \hspace{0.3in} 0 \hspace{0.3in} 0 \hspace{0.3in} 0 \hspace{0.3in} 0 \hspace{0.3in} 0 \hspace{0.3in} 0 \hspace{0.3in} 0 \hspace{0.3in} 0\\
   0  \hspace{0.3in} 0 \hspace{0.3in} 0 \hspace{0.3in} 0 \hspace{0.3in} 0 \hspace{0.3in} 0 \hspace{0.3in} 0 \hspace{0.3in} 0 \hspace{0.3in} 0 \hspace{0.3in} 0\\
   0  \hspace{0.3in} 0 \hspace{0.3in} 0 \hspace{0.3in} 0 \hspace{0.3in} 0 \hspace{0.3in} 0 \hspace{0.3in} 0 \hspace{0.3in} 0 \hspace{0.3in} 0 \hspace{0.3in} 0\nonumber\\
   \end{array}
   \right ),
\end{eqnarray}
\begin{eqnarray}
\beta_{4}= \left ( \begin{array}{llllllllll}
   0  \hspace{0.3in} 0 \hspace{0.3in} 0 \hspace{0.3in} 1 \hspace{0.3in} 0 \hspace{0.3in} 0 \hspace{0.3in} 0 \hspace{0.3in} 0 \hspace{0.3in} 0 \hspace{0.3in} 0\\
   0  \hspace{0.3in} 0 \hspace{0.3in} 0 \hspace{0.3in} 0 \hspace{0.3in} 0 \hspace{0.3in} 0 \hspace{0.3in} 1 \hspace{0.3in} 0 \hspace{0.3in} 0 \hspace{0.3in} 0\\
   0  \hspace{0.3in} 0 \hspace{0.3in} 0 \hspace{0.3in} 0 \hspace{0.3in} 0 \hspace{0.3in} 0 \hspace{0.3in} 0 \hspace{0.3in} 0 \hspace{0.3in} 1 \hspace{0.3in} 0\\
   0  \hspace{0.3in} 0 \hspace{0.3in} 0 \hspace{0.3in} 0 \hspace{0.3in} 0 \hspace{0.3in} 0 \hspace{0.3in} 0 \hspace{0.3in} 0 \hspace{0.3in} 0 \hspace{0.3in} 1\\
   0  \hspace{0.3in} 0 \hspace{0.3in} 0 \hspace{0.3in} 0 \hspace{0.3in} 0 \hspace{0.3in} 0 \hspace{0.3in} 0 \hspace{0.3in} 0 \hspace{0.3in} 0 \hspace{0.3in} 0\\
   0  \hspace{0.3in} 0 \hspace{0.3in} 0 \hspace{0.3in} 0 \hspace{0.3in} 0 \hspace{0.3in} 0 \hspace{0.3in} 0 \hspace{0.3in} 0 \hspace{0.3in} 0 \hspace{0.3in} 0\\
   0  \hspace{0.3in} 0 \hspace{0.3in} 0 \hspace{0.3in} 0 \hspace{0.3in} 0 \hspace{0.3in} 0 \hspace{0.3in} 0 \hspace{0.3in} 0 \hspace{0.3in} 0 \hspace{0.3in} 0\\
   0  \hspace{0.3in} 0 \hspace{0.3in} 0 \hspace{0.3in} 0 \hspace{0.3in} 0 \hspace{0.3in} 0 \hspace{0.3in} 0 \hspace{0.3in} 0 \hspace{0.3in} 0 \hspace{0.3in} 0\\
   0  \hspace{0.3in} 0 \hspace{0.3in} 0 \hspace{0.3in} 0 \hspace{0.3in} 0 \hspace{0.3in} 0 \hspace{0.3in} 0 \hspace{0.3in} 0 \hspace{0.3in} 0 \hspace{0.3in} 0\\
   0  \hspace{0.3in} 0 \hspace{0.3in} 0 \hspace{0.3in} 0 \hspace{0.3in} 0 \hspace{0.3in} 0 \hspace{0.3in} 0 \hspace{0.3in} 0 \hspace{0.3in} 0 \hspace{0.3in} 0
   \end{array}
   \right ).
\end{eqnarray}
When $F_{\mu\nu}$ are taken as antisymmetry and symmetry tensors,
we obtain the two kinds of spinor wave equations (20) and (24),
which maybe describe the left and right circularly polarized
photons. \vskip 8pt

{\bf 3. Spinor wave equation of photon in medium}
\vskip 8pt

In section 2, we give the spinor wave equation of free photon, but
it is more important to study the characteristic of photon in
medium, and it is essential to give the spinor wave equation of
photon in medium.

In Eq. (20) or (24), the spinor wave equation of free photon can
be written as
\begin{equation}
(\beta_{4}i\hbar\frac{\partial}{\partial{t}}+ic\cdot
i\hbar\vec{\beta}\cdot\vec{\nabla})\psi=0,
\end{equation}
For a free photon, the relation between $E$ and momentum $p$ is
\begin{equation}
E=cp,
\end{equation}
the spinor wave equation (27) is obtained by
quantizing equation (28).\\
For a unfree photon, the relation between $E$ and momentum $p$ is
\begin{equation}
E-V=cp
\end{equation}
where $V$ is the interaction potential between photon and medium.

Comparing  equation (28) with (29), and with the aid of equation
(27), we can obtain the quantized equation of (29), it is
\begin{equation}
(\beta_{4}(i\hbar\frac{\partial}{\partial{t}}-V)+ic\cdot
i\hbar\vec{\beta}\cdot\vec{\nabla})\psi=0.
\end{equation}
Eq. (30) is the spinor wave equation of photon in medium.

The interaction potential between photon and medium is [3]
\begin{equation}
V=\frac{hc}{\lambda}(\frac{1}{n}-1)=h\nu(1-n),
\end{equation}
where $\nu$ is photon frequency, and $n$ is medium refractive
indexes.

Substituting Eq. (31) into (30), there is
\begin{equation}
(\beta_{\mu}\partial_{\mu}+\frac{2\pi}{\lambda}\beta_{4}
(1-n))\psi=0.
\end{equation}
When refractive indexes $n=1$, Eq. (32) becomes free photon
equation (22) or (24).

\vskip 8pt

{\bf 4. The covarience of spinor wave equation}

\vskip 8pt

The relativistic quantum theory and quantum field theory should be
covariant, the spinor wave equations (20) and (24) should be also.

At Lorentz transformation
\begin{equation}
x'_{\mu}=\alpha_{\mu\nu}x_{\nu},
\end{equation}
to get
\begin{equation}
x_{\mu}=\alpha_{\nu\mu}x'_{\nu}
\end{equation}
and
\begin{equation}
\partial_{\mu}=\alpha_{\nu\mu}\partial'_{\nu},
\end{equation}
substituting Eq. (35) into (20) or (24), there is
\begin{equation}
\alpha_{\nu\mu}\beta_{\mu}\cdot\partial'_{\nu}\psi=0,
\end{equation}
defining
\begin{equation}
\alpha_{\nu\mu}\beta_{\mu}=L^{-1}\beta_{\nu}L,
\end{equation}
i.e.,
\begin{equation}
L\alpha_{\nu\mu}\beta_{\mu}L^{-1}=\beta_{\nu}.
\end{equation}
Eq. (36) becomes
\begin{equation}
L^{-1}\beta_{\nu}\partial'_{\nu}L\psi=0,
\end{equation}
defining
\begin{equation}
\psi'(x')=L\psi.
\end{equation}
Eq. (39) becomes
\begin{equation}
\beta_{\nu}\partial'_{\nu}\psi'(x')=0.
\end{equation}
The covarience of Eqs. (20) and (24) are proved, and we can give
the transformation $L$.

For a infinitesimal Lorentz transformation
\begin{equation}
x_{\mu}\rightarrow
x'_{\mu}=(\delta_{\mu\nu}+\varepsilon_{\mu\nu})x_{\nu},
\end{equation}
where
\begin{equation}
\varepsilon_{\mu\nu}=-\varepsilon_{\nu\mu}, \hspace{0.5in}
|\varepsilon_{\mu\nu}|<<1,
\end{equation}
writing Eq. (42) with infinitesimal operator, it is
\begin{equation}
x'_{\mu}=(\delta_{\mu\nu}+\frac{1}{2}\varepsilon_{\rho\sigma}I_{\rho\sigma}^{\mu\nu})x_{\nu},
\end{equation}
comparing equation (42) with (44), there is
\begin{equation}
\frac{1}{2}\varepsilon_{\rho\sigma}I_{\rho\sigma}^{\mu\nu}=\varepsilon_{\mu\nu},
\end{equation}
under the infinitesimal Lorentz transformation, the spinor
infinitesimal transformation is
\begin{equation}
\psi'(x')=(1+\frac{1}{2}\varepsilon_{\rho\sigma}I_{\rho\sigma})\psi(x)
\end{equation}
where $I_{\rho\sigma}$ is the infinitesimal operator (matrix) of
Lorentz group.

Comparing equation (40) with (46), we obtain the transformation
matrix $L$
\begin{equation}
L=1+\frac{1}{2}\varepsilon_{\rho\sigma}I_{\rho\sigma}
\end{equation}
and
\begin{equation}
L^{-1}=1-\frac{1}{2}\varepsilon_{\rho\sigma}I_{\rho\sigma},
\end{equation}
substituting Eqs. (47) and (48 )into (38), there is
\begin{equation}
(1+\frac{1}{2}\varepsilon_{\rho\sigma}I_{\rho\sigma})(\delta_{\nu\mu}+\varepsilon_{\nu\mu})\beta_{\mu}
(1-\frac{1}{2}\varepsilon_{\rho\sigma}I_{\rho\sigma})=\beta_{\nu},
\end{equation}
expanding Eq. (50) to one order of $\varepsilon$, we obtain
\begin{equation}
\frac{1}{2}\varepsilon_{\rho\sigma}(I_{\rho\sigma}\beta_{\nu}-\beta_{\nu}I_{\rho\sigma})=-\varepsilon_{\nu\mu}\beta_{\mu}.
\end{equation}
Eq. (50) give the relation between the infinitesimal operator
$I_{\rho\sigma}$ and matrix $\beta_{\nu}$, i.e., the
transformation $L$ is existential. The spinor wave equationS (20)
and(24) are covariant.

\newpage

{\bf 5. Conclusion} \vskip 5pt

In classical electromagnetism theory, the 4-vector potential
$A_{\mu}$ satisfies the differential equation of space-time two
order. In this paper, we give the spinor wave equations of free
and unfree photon, which are the differential equation of
space-time one order, and we prove the spinor wave equations of
free photon is covariant. For the unfree photon, the spinor wave
equations can be used to studied the interaction between photon
and medium, photonic crystals and so on.

\end{document}